# Lightning Ball (Ball Lightning) Created by Thunder, Shock-Wave

Domokos TAR
*M. of Physics ETH-Zürich*
*Eichtlenstr. 16  CH-8712 Stäfa*
*http://www.balllightningtheory.ch*
info@balllightningtheory.ch

Following his observation the author has described in the papers [1] to [4] the formation of a Lightning Ball (LB) with the help of a new theory of symmetry breaking of the vortex ring. In the frame of this theory he emphasizes the primordial rule of the thunder (shock-wave) during the lightning strike in the creation of the LB. The shock-waves and the sound waves propagate very fast but the subsequent enlargement of the vortex ring (air masses) is very slow. This is the reason why the rotating air ring (cylinder) only appeared some seconds later in the observation because of the great inertia of the system. As a consequence of the stationary Mach shock-waves reflections theory the stable distance of the LB to the ground and its very stable horizontal path, in spite of strong winds and rain can, be explained.

## 1. Introduction: Description of a common lightning strike

In the work of Rakov and Uman [5] most properties of a common lightning strike are well summarized. Here is a description of only the relevant details.
The first flash is mostly a negative discharge from cloud to ground (dart leader). After that several return strokes happen in a very short time from the ground to the cloud [6], with intensities greater than for the dart leader. The common speed of these flashes is about one third the speed of light. It has been observed that each channel is tortuous, that means each channel is composed of different short cylindrical segments (straight line) of a length of about 5 to 70 m ([7], [16]).
The ionisation channel is heated up in about 30 microseconds to about 30 thousand degrees and a high overpressure to 10 bar) [9], [6]. Since the channel diameter cannot expand quickly enough the expansion causes a cylindrical shock-wave at about 10 mach (supersonic) velocity [11[, [8]. This phenomenon happens several times throughout the thunder (about 0.3 seconds). Shortly after the overpressure the under pressure (rarefaction) comes (fig.1 and 2). The whole phenomenon is called the N- wave pressure [10] because it resembles a capital N (fig. 1). Fig. 2 shows a typical over and underpressure measured above ground level near to the ground. In the author's observation the rotating direction of the cylindrical ring proves this (fig.1 in [3]). Flying objects with supersonic velocities include explosions, volcanic-eruptions, bullets, missiles, airplanes, satellites and meteors. Szabo I. has proven that the tip of a whip also gives a supersonic crack [8].

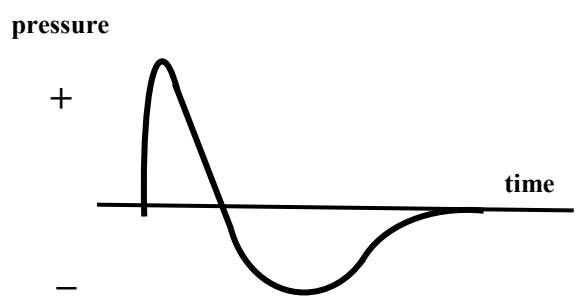

**Fig. 1: Pressure course of a Shock N-Wave**

It is assumed, that the speed of the shock-wave attenuates in the neighbourhood of the channel at a distance of about 6 times the channel's diameter [9] and progresses further only with normal velocity. Thus the Mach number (ratio between supersonic/sound velocities) is very time dependent [13]. It is assumed that at the ground point a half-spherical shock wave dominates and the whole lightning can theoretically be simply represented by a spherical shock-wave [12], [14], [11]. Few A. considered the whole lightning flash as a single element of radiator. After the shock wave the thunder transforms in a normal sound radiation. The observer hears a superposition of many shock waves and sounds. This ranges in frequency between 1 and 600 Hz. There is a component under 20 Hz of infrasound inaudible to humans. It is believed that this infrasound is also created by the change of the clouds electrostatic charges [27].

The shock-wave with its N-shaped form creates a turbulent vortex ring in the simplest form [10]. Similar vortex rings have been observed in different hydrodynamic flows: the photographs of Okabe, J. and Inoue, S. fig.3, [17], Kopiev V.F. fig.4, [18], Crum L.A. fig.5, [19], [20], Glezer, A. fig.6, [21] and "Smoke ring" on Wikipedia fig.7, [22]. L.A. Crum proved that ultrasound waves in water can create a growing vortex ring, the vacuum bubble of which implodes later through the emission of a high intensity shock-wave.

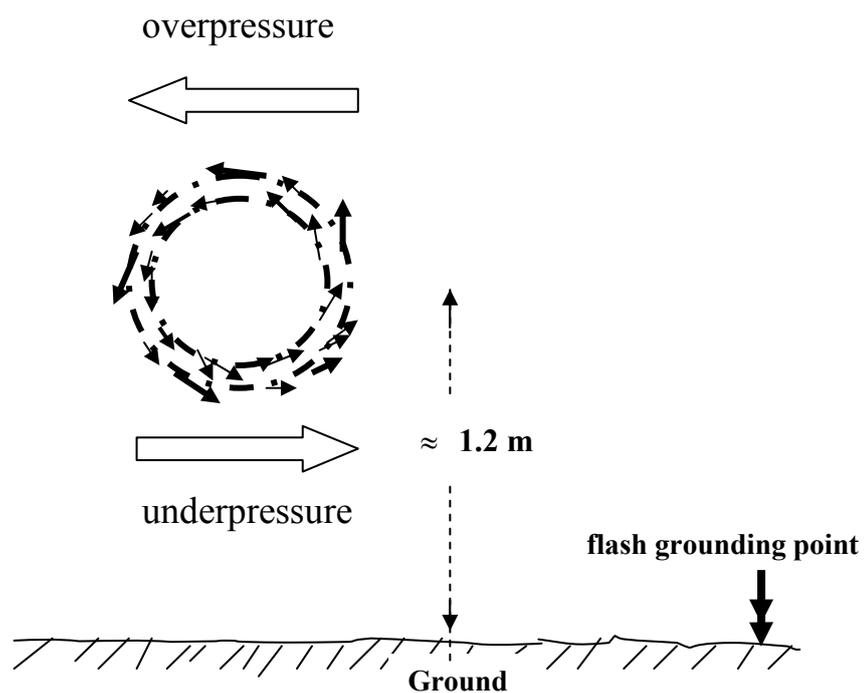

**Fig.2**: **Rotating air-ring (cylinder) created by a shock wave**

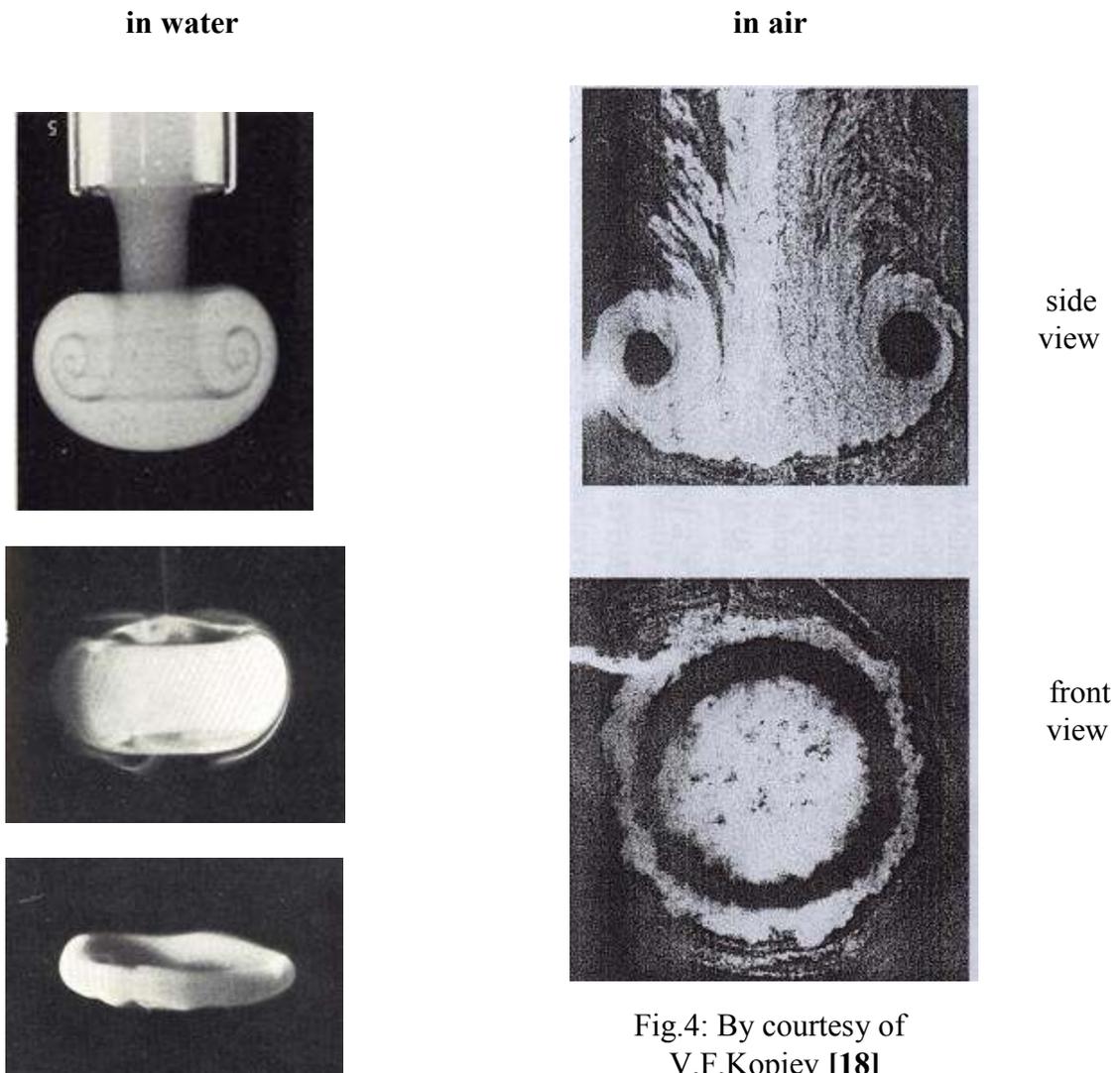

**Fig.3 and 4: Different stages of vortex rings [17], [18]**

## 2. Theory of the Lightning Ball formation

The author supposes, that the lightning strike, thunder and the shock-wave create first an invisible hydrodynamic vortex ring that moves horizontally at low speed. The rotating speed may be high, but critically the rotation is only in the poloidal plane (in the surface of cross section). There is no rotation in the toroidal direction at all (see the mentioned figures).The Lightning Ball may possibly been created by symmetry breaking of this vortex ring [1], [2], [3]. This insight in to the vortex torus physics could help in the Tokamak etc. of nuclear fusion torus experiments [4]. The photographs mentioned before have been taken with high speed photography. A. Glezer [21] photographed in the cross sectional view (fig. 6).  H.L. Reed shows in her paper [23] a gallery of vortex rings.

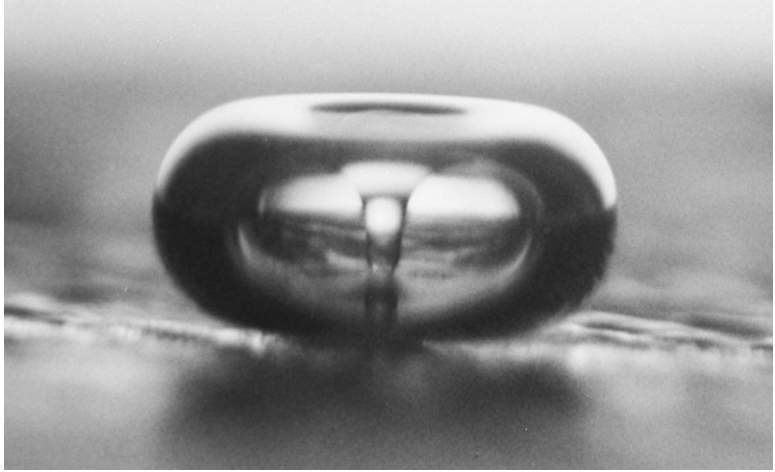

**Fig.5**: **High speed photograph of water vortex vacuum bubble collapse**
by courtesy of L. Crum [19], [20].

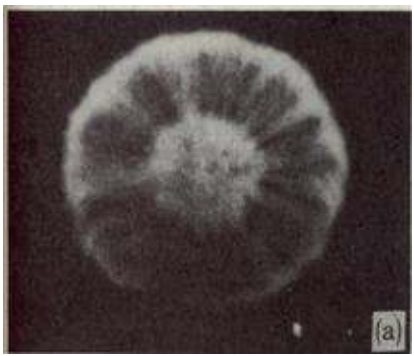

**Fig.6: High speed photograph
of water vortex ring**
by courtesy of A. Glezer [21]

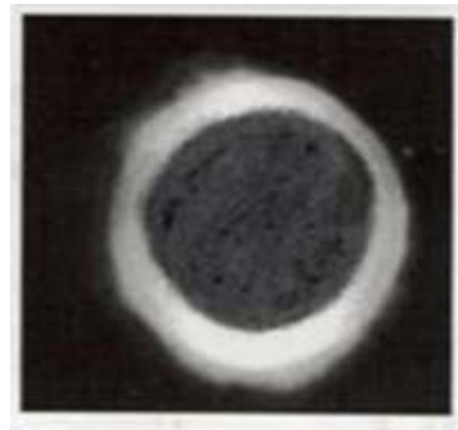

**Fig.7: Smoke ring**
(Wikipedia [22])

In the observation there was a single bush in the vicinity of the lightning strike [1]. This bush probably cut the expanding vortex ring, without the bush the ring would have expanded and dissipated. Also the bush possibly cut the ring that in turn suddenly shrank itself into a ball, following the law of conservation of mass, momentum and energy. First of all a rotating ring was observed [3] and only in the last phase of its rotation did the Lightning Ball appear. At the same time as the turning ring disappeared. A certain time was needed until the ball appeared, this means that the visible light emission was due to the triboluminescence [1[, [2] and electroluminescence phenomenon [25]. The turning of the cylinder endured for about 5 seconds. At first the author thought, with great surprise, that he saw a kind of tornado, but with a horizontal axis.

## 3. The rule of the thunder, shock-wave in the creation of the Lightning Ball

It has been proved by means of high speed photography ( 4-to 10 thousand pictures per second) [26] of objects at supersonic velocities, that there are different types of shock-wave reflections depending on the reflecting surface angle to the shock wave direction [15]. Courant and Friedrichs (1948) indicated theoretically that there are possibly three different kinds of Mach Reflection wave configurations. Ben-Dor & Takayama (1986/7) validated this experimentally [15]. In the book of G.Ben-Dor [15] the reader can find all kinds of reflections and theories for the Mach reflections. Here are the very strange and uncommon reflection configurations and laws. The physical reason for this is that a shock-wave "knows" well in advance that an obstacle is there and it "adjusts" itself before it reaches the reflecting surface. This means it can "negotiate" (at the triple point) with the obstacle beforehand and it blocks its proper propagation path and changes direction in advance (fig.8).

One kind of Mach reflection pattern called the Stationary Mach Reflection (StMR) is where the triple point moves parallel to the reflecting surface and is valid for unsteady flows. In this case the reflecting surface is the ground where the lightning happened (fig.8). For better understanding in this field further research is needed. The distance from the triple point to the reflecting surface is called the stem. This can be a different length according to the actual situation. In the LB observation the ball was about 1.2 meters above the ground.

Fig. 8 shows us this type of Machs reflection (which is axisymmetrical in space), that the author thinks corresponds well to his LB observation. This is when the triple point moves parallel to the reflecting surface after the turbulent vortex ring has been created by the shock-wave. The StMR is valid for non steady flows, as was the case.

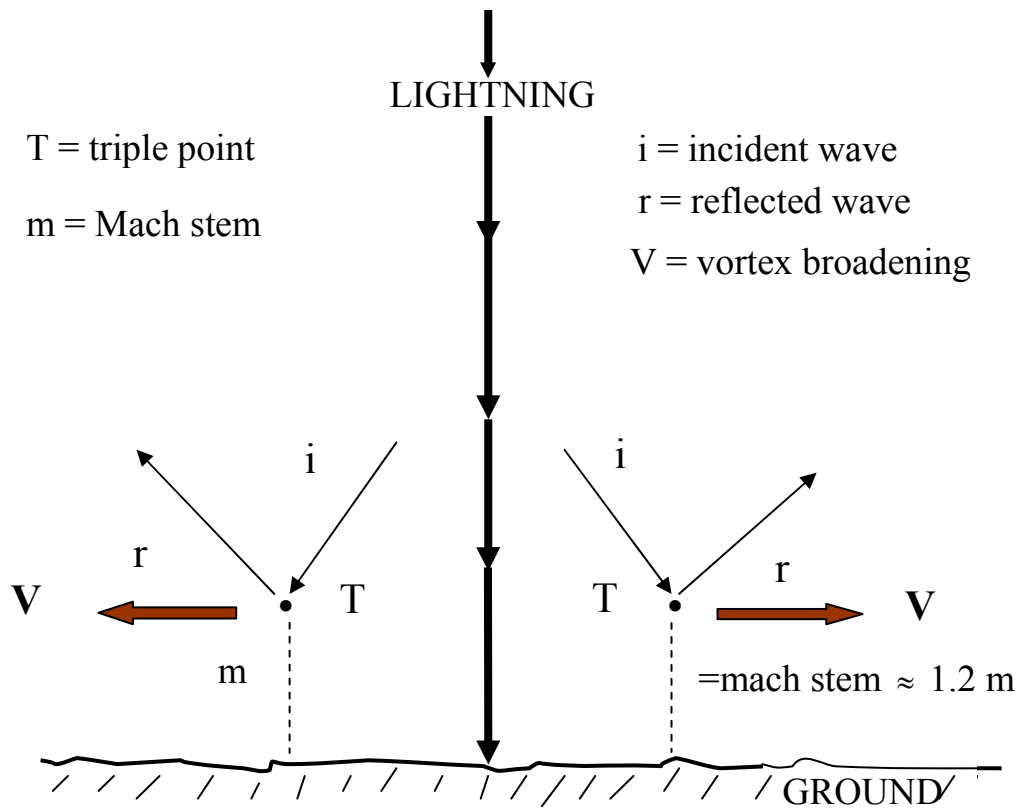

**Fig.8: Air vortex ring created by reflected shock wave [15]**
by courtesy of Gabi Ben-Dor, (modified)

The LB, after its creation, moved absolutely parallel to the ground in spite of the strong turbulence. This indicated that the LB preserved its parallel path to the ground after the creation of the vortex ring.

**Conclusion:**

As we have seen many points of this theoretical description are consistent with the observation.

## 4. Comments on two Ball Lightning observations

The author believes, that LB and Ball Lightning (BL) are two different phenomena. Here he depicts two BL observations with photographs and he thinks these are true BL's:

The first example: Stekol'nikov S. showed in his article: *"Thunderstorm at Black Sea"* a photograph of a presupposed BL [28]. The same photograph can be seen in the paper of A.I. Nikitin et al [29].

The second presupposed BL photograph is shown on the internet, taken by a student in Nagano (Japan) in 1987 [30].

The reason for the assumption of true BL is, that both BL photographs have the same characteristics as Gallimberti et al have shown in their paper [24].